\documentclass[aps,prb,twocolumn,floatfix,superscriptaddress,showpacs]{revtex4}

\usepackage{graphicx}
\usepackage{bbm}
\usepackage{amsmath,amssymb}
\newcommand{\be}{\begin{equation}}
\newcommand{\beq}{\begin{equation}}
\newcommand{\ee}{\end{equation}}
\newcommand{\bea}{\begin{eqnarray}}
\newcommand{\eea}{\end{eqnarray}}
\newcommand{\ba}{\begin{array}}
\newcommand{\ea}{\end{array}}

\renewcommand{\vr} {{\bf r}}
\newcommand{\vj} {{\bf j}}

\def\v#1{\mbox{\boldmath $#1$}}

\renewcommand{\vr} {{\bf r}}
\newcommand{\nn}{\nonumber}

\begin{document}
\title{Becke-Johnson-type exchange potential for two-dimensional systems}
\author{S. Pittalis}
\email[Electronic address:\;]{pittaliss@missouri.edu}
\affiliation{Institut f{\"u}r Theoretische Physik,
Freie Universit{\"a}t Berlin, Arnimallee 14, D-14195 Berlin, Germany}
\affiliation{European Theoretical Spectroscopy Facility (ETSF)}
\affiliation{Department of Physics and Astronomy, University of Missouri, Columbia, Missouri 65211, USA}
\author{E. R{\"a}s{\"a}nen}
\email[Electronic address:\;]{erasanen@jyu.fi}
\affiliation{Nanoscience Center, Department of Physics, University of
  Jyv\"askyl\"a, FI-40014 Jyv\"askyl\"a, Finland}
\author{C. R. Proetto}
\altaffiliation[Permanent address: ]{Centro At{\'o}mico Bariloche and Instituto Balseiro, 8400
S.C. de Bariloche, R{\'i}o Negro, Argentina}
\affiliation{Institut f{\"u}r Theoretische Physik,
Freie Universit{\"a}t Berlin, Arnimallee 14, D-14195 Berlin, Germany}
\affiliation{European Theoretical Spectroscopy Facility (ETSF)}

\date{\today}

\begin{abstract}
We extend the Becke-Johnson approximation [J. Chem. Phys. {\bf 124},
221101 (2006)] of the exchange potential to two dimensions.
We prove and demonstrate that a direct extension of the underlying
formalism may lead to divergent behavior of the
potential. We derive a cure to the
approach by enforcing the gauge invariance and correct asymptotic
behavior of the exchange potential. The procedure leads to an
approximation which is shown, in various quasi-two-dimensional
test systems, to be very accurate 
in comparison with the exact exchange potential, and thus a 
considerable improvement over the commonly applied 
local-density approximation.
\end{abstract}

\pacs{31.15.E-; 71.15.Mb; 73.21.La}

\maketitle

\section{Introduction}

The advent of density-functional theory~\cite{dft1,dft2} 
(DFT) was followed by the development of approximate functionals 
for the exchange and correlation energy of many-electron systems.
Significant advances were achieved beyond the local-density 
approximation (LDA) by generalized-gradient approximations, orbital 
functionals, and hybrid functionals.~\cite{dft3} Those efforts focused almost
solely on three-dimensional systems. However, already since the 
1970s systems of reduced dimensionality became of great relevance in
solid state physics. The present nanodevices consist of
a large variety of low-dimensional systems, where the many-body
effects of interacting electrons need to be addressed. Of particular 
interest are two-dimensional (2D) structures including, e.g.,
semiconductor layers and surfaces, quantum Hall systems, and various 
types of quantum dots.~\cite{qd} 

Previous studies have shown
that density functionals developed particularly for 3D fail when
applied to quasi-2D systems.~\cite{cost1,kim,pollack}
By ``quasi-2D'' we mean physical systems whose main
building block is the {\em quasi-2D electron gas}. This class
comprises most systems listed
in the previous paragraph and they can be reliably treated by a
``pure'' 2D approach, i.e., on a 2D grid. The physical reason 
for this is that the confinement potential along
the growth direction (say $z$), is typically much stronger than in the two
other spatial dimensions. This results in a quantum mechanical supression
of the degrees of freedom along the $z$ direction, and the system becomes
effectively 2D. Then, the influence of the 
surrounding host material is dealt within the effective-mass 
approximation manifesting itself as an effective
mass and a dielectric constant in the 2D Hamiltonian.~\cite{qd}

Within the DFT approach, 2D many-electron problems are most 
commonly treated using
the 2D-LDA exchange functional~\cite{rajagopal} combined with the 2D-LDA
correlation parametrized first by Tanatar and Ceperley~\cite{tanatar}
and later, for the complete range of collinear spin polarization, by
Attaccalite and co-workers.~\cite{attaccalite} Despite the relatively
good performance of LDA with respect to, e.g., quantum Monte Carlo
calculations~\cite{henri} in semiconductor quantum dots,
there is a clear lack of accurate 2D density functionals to deal with
diverse situations, especially in the strongly correlated regime. 
Only recently, the transition from 3D to 2D has
been analyzed,~\cite{cost2} and exchange-correlation {\em energy} functionals
specially tailored for the 2D world have been 
proposed.~\cite{x1,ring,nicole,lccs,dga,correlation,gamma,rp}
Also very recently, a DFT framework specially tailored for strongly
correlated electrons has been developed and successfully 
tested for few-electron quantum dots.~\cite{paola}

In the present work, we aim at moving another step along the direction
from 3D to 2D. We focus on approximations for the exchange {\em
  potential}, where our aim is to achieve an accuracy comparable to 
the exact result -- available from the optimized-effective-potential (OEP) 
method~\cite{SharpHorton:53,TalmanShadwick:76} or its
approximations such as the Krieger-Lie-Iafrate (KLI)
potential~\cite{KriegerLiIafrate:92} -- yet simplifying the 
computational effort. We are able to find such an approximation by (i)
extending the framework of the well-known Becke-Johnson 
potential~\cite{BJ} to 2D, and (ii) requiring the potential to satisfy
the gauge invariance and correct asymptotic behavior. 
Our spin- and current-dependent exchange potential is indeed 
well comparable to the KLI and superior to the LDA in 2D quantum dots 
with varying sizes and varying external magnetic fields.

\section{Formalism}

\subsection{Spin-density functional theory}

Considering open-shell systems, (collinear) spin-DFT (SDFT) 
formalism~\cite{Barth:79} is one of the most
useful. In this case, the total energy $E$ of a 2D system of 
interacting electrons is a functional of the two spin densities,
$\rho_{\sigma}(\vr)$ with  $\sigma=\uparrow,\downarrow$:
\bea\label{ETOT}
\lefteqn{
E[\rho_{\uparrow},\rho_{\downarrow}] = 
T_{s}[\rho_{\uparrow},\rho_{\downarrow}] } \nn \\
&& + \int d^2 r  \; v_{0\sigma}(\vr) \rho(\vr)
+ U[\rho] + E_{xc}[\rho_{\uparrow},\rho_{\downarrow}] \; ,
\label{etot-sdft}
\eea
where
\be
T_{s}[\rho_{\uparrow},\rho_{\downarrow}] = 
\sum_{\sigma = \uparrow,\downarrow} \sum_{j}^{N_{\sigma}} 
\int d^2 r  \; \varphi_{j \sigma}^*(\vr) \left( - \frac{\nabla^2}{2} \right) 
\varphi_{j \sigma}(\vr) \; ,
\label{ts}
\ee
is the non-interacting kinetic energy, the sum runs over occupied states, 
and $N_{\sigma}$ is the number of 
electrons with spin $\sigma$. Hartree atomic units (a.u.) are used, unless
stated otherwise. Above, $v_{0 \sigma}(\vr)$ is the spin-dependent external 
potential. 
The classical electrostatic (Hartree) interaction energy is given by 
\be
U[\rho] = \frac{1}{2} \int d^2 r  \int d^2 r' \; \frac{\rho(\vr) \rho(\vr')}{| \vr - \vr'|} \; ,
\ee
where 
\be
\rho(\vr) = \rho_{\uparrow}(\vr) + \rho_{\downarrow}(\vr)
\label{rho}
\ee
is the total electronic density, which is the sum of the spin densities
\be
\rho_{\sigma}(\vr) = \sum_{j=1}^{N_{\sigma}} | \varphi_{j \sigma}(\vr) |^2 \; ,
\label{rhosigma}
\ee
with the sum running over occupied orbitals.
Besides, 
$E_{xc}[\rho_{\uparrow},\rho_{\downarrow}]$ is the exchange-correlation energy functional,
which in practice needs to be approximated. The single-particle orbitals 
$\varphi_{j \sigma}(\vr)$ in Eq.~(\ref{ts}) are solutions of the Kohn-Sham (KS)
equation~\cite{BarthHedin:72} 
\be 
\left( -\frac{\nabla^2}{2} + v_{s\sigma}(\vr) \right) \varphi_{j \sigma}(\vr) 
= \varepsilon_{j \sigma} \varphi_{j \sigma}(\vr) \; ,
\label{kseq}
\ee
where $j$ is a collective index for the one-electron quantum numbers, except 
for the spin. The effective single particle potential for spin $\sigma$ is given by 
\be
v_{s\sigma}(\vr) = v_{0\sigma}(\vr) + v_{\rm H}(\vr) + v_{xc\sigma}(\vr) \; ,
\label{espp}
\ee
with the Hartree potential
\be
v_{\rm H}(\vr) = \int d^2 r' \frac{\rho(\vr')}{| \vr - \vr'|} \; ,
\ee
and the exchange-correlation potential 
\be
v_{xc\sigma}(\vr) = \frac{\delta E_{xc}[\rho_{\uparrow},\rho_{\downarrow}]}
{\delta\rho_{\sigma}(\vr)} \; .
\label{fdexc}
\ee

\subsection{Optimized-effective-potential method}

In the following, we restrict ourselves to orbital functionals.
A standard example of this type of functionals is the
exchange energy functional of the form
\begin{widetext}
\be
E_{x}[\rho_{\uparrow},\rho{\downarrow}] = -\frac{1}{2} \sum_{\sigma=\uparrow,\downarrow} \sum_{j,k=1}^{N_\sigma} 
\int d^2 r \int d^2 r' \frac{\varphi_{j \sigma}^*(\vr)\varphi_{k \sigma}^*(\vr')
\varphi_{j \sigma}(\vr')\varphi_{k \sigma}(\vr)}{|\vr
- \vr'|} \; ,
\label{Ex}
\ee
\end{widetext}
which resembles the Fock term of Hartree-Fock theory, here evaluated 
with KS orbitals. This is the so-called
exact-exchange (EXX) energy functional. It should be noted that $E_{x}[\rho_{\uparrow},\rho{\downarrow}]$
is an implicit functional of the spin densities $\rho_{\sigma}(\vr)$. 
Of course, correlation can be considered
as well in a similar framework.
In this kind of functionals the calculation of the exchange-correlation potential 
requires the OEP method.~\cite{SharpHorton:53,TalmanShadwick:76} 
For a review of the method the reader is referred to 
Refs.~\onlinecite{KummelKronik,Engel:03,GraboKreibichKurthGross:00}. 
Restricting ourselves to the EXX case, the OEP method leads to an integral equation for the exchange 
potential, which can be written in compact form as
\be
\sum_{j=1}^{N_{\sigma}} \left( \psi_{j \sigma}^*(\vr) \varphi_{j \sigma}(\vr) 
+ c.c. \right) = 0,
\label{oep}
\ee
where the so-called orbital shifts are defined as
\be\label{oshift}
\psi_{j \sigma}^*(\vr) = \int d^2 r ' \varphi^*_{j \sigma}(\vr')
\left[ 
v_{x\sigma}(\vr')-u_{x j\sigma}(\vr') \right] 
G_{j \sigma}^{S}(\vr',\vr).
\ee
Here $G_{j \sigma}^S$ is the Green function of the KS system,
\be\label{GKS}
G_{j \sigma}^{S}(\vr',\vr) = \sum_{\stackrel{k=1}{\varepsilon_{k \sigma} \neq 
\varepsilon_{j \sigma}} }^{\infty}
\frac{\varphi_{k \sigma}^*(\vr')\varphi_{k \sigma}(\vr)}
{\varepsilon_{j \sigma}-\varepsilon_{k \sigma}} \; ,
\ee
and
\be\label{uxsi}
u_{x j \sigma}(\vr) = 
\frac{1}{\varphi_{j \sigma}^*(\vr)}
\frac{\delta E_{x}}{\delta \varphi_{j \sigma}(\vr)} \; .
\ee

In a series of steps,\cite{KriegerLiIafrate:92-2,GraboKreibichKurthGross:00}
the OEP equation as given in Eq.~(\ref{oep}) can be transformed to
\begin{widetext}
\be
v_{x\sigma}(\vr) =
\frac{1}{2 \rho_{\sigma}(\vr)} 
\sum_{j=1}^{N_\sigma} \bigg[ |\varphi_{j \sigma}(\vr)|^2 
\left( u_{x j\sigma}(\vr) + 
\left( \bar{v}_{x j\sigma}-\bar{u}_{x j\sigma} \right) 
\right) -\nabla \cdot 
( \psi_{j \sigma}^*(\vr) \nabla \varphi_{j \sigma}(\vr) ) 
\bigg] +c.c. \; ,
\label{oep-sdft}
\ee
\end{widetext}
where
\be\label{vbar}
\bar{v}_{x j\sigma} =  \int d^2 r  \; 
\varphi_{j \sigma}^*(\vr) v_{x \sigma}(\vr) \varphi_{j \sigma}(\vr) \;,
\ee
and
\be\label{ubar}
\bar{u}_{x j\sigma} =  \int d^2 r  \; 
\varphi_{j \sigma}^*(\vr) u_{x j\sigma}(\vr) \varphi_{j \sigma}(\vr) \; .
\ee
The OEP equations can be solved iteratively, simultaneously with
the corresponding KS equations, in a self-consistent fashion.

Due to the presence of the unoccupied KS orbitals
in the definition of the orbital shifts [see Eqs.~(\ref{oshift}) and 
(\ref{GKS})], the full numerical solution of the OEP integral equation is 
nontrivial. Of course, one may take advantage by specifying it for a particular kind of systems. 
In the original paper,~\cite{TalmanShadwick:76} solutions were 
presented for atomic systems with spherical symmetry. Much later, the
OEP equation has also been solved for systems with lower symmetry such as molecules, 
\cite{IvanovHirataBartlett:99,Goerling:99} 
solids,\cite{StaedeleMajewskiVoglGoerling:97} metallic surfaces,\cite{hrp} and quasi two-dimensional
electron gases at the interface of two different semiconductors.\cite{rp} 
In addition, an iterative algorithm for the solution of the OEP equation based on the orbital shifts 
has been implemented.\cite{KuemmelPerdew:03,KuemmelPerdew:03-2} 
Recently, progress has been made in studying physically relevant
examples of non-collinear magnetism,\cite{Sharma1,Sharma2} and open-shell systems
in the relativistic limit.\cite{demo1,demo2}
On the other hand, one may approximate the full OEP in order to 
save computational effort. Along this line, the KLI
approach~\cite{KriegerLiIafrate:92} has turned out to be rather accurate in many 
situations. In the KLI approximation, the terms containing the orbital shifts on the r.h.s. of 
Eq.~(\ref{oep-sdft}) are neglected completely.\cite{GraboKreibichKurthGross:00}

For the analysis which follows in the next sections, we may 
rewrite Eq.~(\ref{oep-sdft}) as
\bea
v_{x \sigma}(\vr) &=& v^{\text{SL}}_{x \sigma}(\vr) + \Delta v^{\text{KLI}}_{x \sigma}(\vr) + \Delta v^{\text{OS}}_{x \sigma}(\vr) 
\; , \nn \\
&=&  v^{\text{SL}}_{x \sigma}(\vr) + \Delta v^{\text{OEP}}_{x \sigma} (\vr) \; ,
\label{vx}
\eea
where
\be
v^{\text{SL}}_{x \sigma}(\vr) =
\frac{1}{2 \rho_{\sigma}(\vr)} 
\sum_{j=1}^{N_\sigma}  |\varphi_{j \sigma}(\vr)|^2 
u_{x j \sigma}(\vr)  +c.c.
\label{vsl}
\ee
is the so-called Slater potential;
\be
\Delta v^{\text{KLI}}_{x \sigma}(\vr) = 
\frac{1}{2 \rho_{\sigma}(\vr)} 
\sum_{j=1}^{N_\sigma} |\varphi_{j \sigma}(\vr)|^2 
\left( \bar{v}_{x j \sigma}-\bar{u}_{x j \sigma} \right)  + c.c.
\label{vkli}
\ee
is the correction to the Slater potential added by the KLI approximation of the
OEP solution; and
\be
 \Delta v^{\text{OS}}_{x \sigma}(\vr) =   
 \frac{-1}{2 \rho_{\sigma}(\vr)} 
\sum_{j=1}^{N_\sigma} |\varphi_{j \sigma}(\vr)|^2 
 \nabla \cdot 
\left( \psi_{j \sigma}^*(\vr) \nabla \varphi_{j \sigma}(\vr) \right) + c.c.
\label{vos}
\ee
is the correction to the KLI solution, which adding the contribution from the
orbital shifts, provide the full OEP solution. Thus, the definition
of $\Delta v^{\text{OEP}}_{x \sigma}(\vr)$ in Eq.~(\ref{vx}) becomes clear.
We remind that the asymptotic limit of $v_{x \sigma}(\vr)$ in Eq.~(\ref{vx}), for finite systems, is given by the first
term, $v_{x \sigma}^{\text{SL}}(r \rightarrow \infty) \rightarrow - \; 1/r$.\cite{GraboKreibichKurthGross:00,ring}

We point out that the derivations above are fully independent 
of the dimensionality. In orbital functionals the dimensional
character is ensured from the structure of the KS orbitals 
themselves, and no further assumptions are made. Therefore, the
OEP (and KLI) schemes presented above apply equally well to 3D and 2D.
Below, however, we focus on a recent approximation to 
$\Delta v^{\text{OEP}}_{x \sigma}$ proposed by Becke and
Johnson,\cite{BJ} (BJ) which is {\em dimension-dependent} by construction.
In 3D, the BJ approximation has received considerable 
attention.~\cite{Armiento,Staroverov,Kodera,Gauduk,Fabien,Kevin,Naoto,Tran}

\section{Becke-Johnson approximation}

\subsection{Three dimensions}

For 3D systems, Becke and Johnson have proposed to 
approximate $\Delta v^{\text{OEP}}_{x \sigma}(\vr)$ by 
$\Delta v^{\text{BJ}}_{x \sigma}(\vr)$,
defined as \cite{BJ}
\be\label{BJ2}
\Delta v^{\text{BJ}}_{x \sigma}(\vr) = \frac{1}{\pi} \left( \frac{5}{12} \right)^{1/2} \sqrt{ \left[ \frac{ \tau_{\sigma}(\vr) }{ \rho_\sigma(\vr) }  \right]},
\ee
with
\begin{equation}\label{3Dtau}
\tau_\sigma(\vr)=\sum_{k=1}^{N_\sigma} |\nabla\varphi_{k \sigma}(\vr)|^2 \; ,
\end{equation}
being (twice) the spin-dependent kinetic-energy density.
This approximation was found by seeking a simple expression 
having the following properties: (i) it is (possibly) invariant with 
respect to unitary orbital transformation; 
(ii) it provides an exact treatment of any (ground state) hydrogenic atom;
(iii) it has a step-like structure characteristic of $\Delta
v^{\text{OEP}}_{x \sigma}(\vr)$ in multi-shell atoms;\cite{KummelKronik,Engel:03,GraboKreibichKurthGross:00}
(iv) it is exact in the 3D uniform electron gas limit, 
$\Delta v^{\text{OEP}}_{x \sigma} = \Delta v^{\text{BJ}}_{x \sigma} = \left[ 3 \rho_{\sigma}/(4\pi) \right]^{1/3}$, 
with constant spin densities.
The BJ potential as defined in Eq.~(\ref{BJ2}) has been a good
approximation for several atoms,~\cite{BJ} and it has given valid
estimations for semiconductor band gaps.~\cite{Fabien,Tran}

\subsection{Extension to two dimensions}

In the framework of the BJ approach, we may suggest the following
approximation to be valid in 2D,
\be\label{BJ2B}
\Delta v_{x \sigma}(\vr)= \frac{2 \sqrt{2}}{3 \pi} \sqrt{ \left[ \frac{ \tau_{\sigma}(\vr) }{ \rho_\sigma (\vr)}  \right]}\;,
\ee
where the coefficient $2\sqrt{2}/(3\pi)$ is chosen to satisfy the exact 2D uniform electron gas limit 
\be
\Delta v^{\text{OEP}}_{x \sigma} = \Delta v_{x \sigma} = 
\frac{4}{3 \sqrt{\pi}} \rho_{\sigma}^{1/2},
\ee
with constant densities.
However, as shown analytically and also numerically in the next section, this
form serves only as the starting point. We need to introduce two
completely general modifications, based on an additional set of physical requirements.

First, we observe that
property (i) given in the previous section is satisfied only under the
restriction of real-valued orbitals. In fact, it is easy to see that 
under gauge transformation for both real and complex-valued orbitals
the BJ contribution to the exchange potential is not gauge 
invariant.
The relevance of the gauge invariance requirement for the
meta-generalized-gradient approximation has been analyzed in some
detail by Tao and Perdew.~\cite{TaoPerdew:05} We expect that the same
criterion should apply to any approximation for $v_{x \sigma}(\vr)$
which depend explicitly on $\tau_{\sigma}(\vr)$.
We can ensure the gauge invariance by replacing 
$\tau_{\sigma}(\vr)$ with~\cite{TaoPerdew:05,gamma}
\be
\label{current}
\tau_{\sigma}(\vr) \longrightarrow \bar{\tau}_{\sigma}(\vr) = \tau_\sigma(\vr) - \frac{\vj^{\;2}_{p \sigma}(\vr)}{\rho_{\sigma} (\vr)},
\ee
where
\begin{equation}
\vj_{p \sigma}(\vr)=\frac{1}{2i}\sum_{k=1}^{N_\sigma} \left[
 \varphi^*_{k \sigma}(\vr) \left(\nabla \varphi_{k \sigma}(\vr)\right) - \left(\nabla \varphi^*_{k \sigma}(\vr)\right) 
\varphi_{k \sigma}(\vr) \right]
\end{equation}
is the spin-dependent paramagnetic current density.

Second, property (ii) given
in the previous section is not valid for {\it all} one-electron systems. As we show
below, this also relates to the wrong (divergent) behavior of the approximation in the
asymptotic region.
As it has been shown in 3D, an ad-hoc solution may
be found \cite{Armiento} for specific cases, but still it would be 
preferable to have a general cure. Hence, 
to obtain an expression which vanishes in the case of a generic single-particle state, we 
add an additional term
\be\label{D}
\bar{\tau}_{\sigma}(\vr) \longrightarrow D_{\sigma}(\vr) = \bar{\tau}_{\sigma}(\vr) - \frac{1}{4}\frac{\left( \nabla \rho_\sigma(\vr)
\right)^2}{\rho_\sigma(\vr)}.
\ee

In conclusion, in 2D we obtain 
\be\label{MBJ}
\Delta v_{x \sigma}(\vr) \longrightarrow \Delta {v}^{\text{C}}_{x \sigma}(\vr) =
 \frac{2 \sqrt{2}}{3 \pi} \sqrt{ \left[ \frac{ D_{\sigma}(\vr) }{ \rho_\sigma(\vr)}  \right]}\;.
\ee
The present approximation for the {\em total} corrected 2D exchange potential reads then
\bea\label{key}
v_{x \sigma}(\vr) & = & v^{\text{SL}}_{x \sigma}(\vr) + \Delta{v}^{\text{C}}_{x \sigma}(\vr) \nonumber \\
& = & v^{\text{SL}}_{x \sigma}(\vr)+\frac{2\sqrt{2}}{3\pi}
\sqrt{ \left[ \frac{ D_{\sigma}(\vr) }{ \rho_\sigma(\vr)}  \right] } \; ,
\eea
with
\be\label{elf}
D_{\sigma}(\vr)= \tau_{\sigma}(\vr)-\frac{1}{4}\frac{\left( \nabla \rho_\sigma(\vr)
\right)^2}{\rho_\sigma(\vr)}-\frac{\vj^2_{p \sigma}(\vr)}{\rho_\sigma (\vr)} \; .
\ee
Eq.~(\ref{key}) is the key expression of the present work, and in the following we will discuss
its properties, first analytically and then numerically. First, it is easy to see that $D_{\sigma} (\vr) \equiv 0$
for {\it any} one-electron system (per spin channel), as it should be [property (i)].
To see this, it is enough to note that for this particular case 
$\rho_{\sigma}(r)= |\varphi_{\sigma}(r)|^2$, with $\varphi_{\sigma}(r)=\sqrt{\rho_{\sigma}(r)}e^{i\theta(r)}$, with
no loss of generality. Substituting this in Eq.~(\ref{elf}), the result $D_{\sigma} (\vr) \equiv 0$ follows at once,
for arbitrary functions $\rho_{\sigma}(\vr)$ and $\theta(\vr)$.
Second, in the asymptotic limit, $D_{\sigma}(r \rightarrow \infty)
\rightarrow 0$, for {\it any} finite system.
To see this, it is enough to note that for any finite system, all ingredients of $D_{\sigma}(\vr)$ in Eq.~(\ref{elf})
become dominated by the contribution of the highest occupied orbital. 
As a consequence, in the asymptotic region the system becomes {\it effectively} one-electron like, and the result
$D_{\sigma}(r \rightarrow \infty) \rightarrow 0$ applies.
This guarantees that $v_{x \sigma}(\vr)$ has the correct asymptotic behavior
given by $v_{x \sigma}^{\text{SL}}(\vr)$.   
Besides correcting the potential in the asymptotic region, we also 
expect improvements at finite $\vr$. We find this reasonable due 
to the fact that $D_{\sigma}(\vr)$ is the main ingredient of the 
so-called electron localization function,\cite{elf1,elf2,elf3} which
is able to deal with situations where the multi-shell structure of atoms becomes relevant, 
as well as in the region of interest for chemical bonding. 

Then, it is reassuring to note that one can justify the choice of the power $\alpha$ in
\be\label{MBJ2}
\Delta {v}^{\text{C}}_{x \sigma}(\vr) =
 \frac{2 \sqrt{2}}{3 \pi} \left[ \frac{ D_{\sigma}(\vr) }{ \rho_\sigma(\vr)}   \right]^\alpha \; ,
\ee
by observing that the resulting potential scales linearly only for
$\alpha=1/2$, similarly to the exact exchange potential.\cite{levyperdew}
Under uniform scaling of the coordinates, $\vr \rightarrow \lambda \vr$ (with $0 < \lambda
< \infty$), and for the norm-preserving many-body wave function, the 2D density scales quadratically with $\lambda$,
$\rho^{\lambda}_{\sigma}(\vr) \rightarrow \lambda^2 \rho_{\sigma}(\lambda \vr)$. This leads to the result that
the KS orbitals in 2D are seen to scale as
$\varphi_{k \sigma}^{\lambda}(\vr) \rightarrow \lambda \; \varphi_{k \sigma}(\lambda \vr)$. Thus,
$\tau_{\sigma}^{\lambda}(\vr) \rightarrow \lambda^4 \; \tau_{\sigma}(\lambda \vr)$,
$\nabla \rho^{\lambda}_{\sigma}(\vr) \rightarrow \lambda^{3}\nabla_{\lambda \vr} \rho_{\sigma}(\lambda \vr)$,
and $\vj^{\lambda}_{p \sigma}(\vr) \rightarrow \lambda^3 \vj_{p \sigma}(\lambda \vr)$.
Substituting these relations to Eq.~(\ref{MBJ2}) yields the scaling relation
\be
\Delta {v}^{\text{C} \; \lambda}_{x \sigma}(\vr) \rightarrow
\lambda^{2 \alpha} \Delta {v}^{\text{C}}_{x \sigma}(\lambda \vr) \; ,
\ee
which fulfills the exact linear scaling constraint {\it only} if $\alpha = 1/2$.

Finally, we remind that the calculation of $\Delta {v}^{\text{C}}_{x
  \sigma}(\vr)$ is computationally cheap in comparison with the full OEP
method or the KLI approximation, which is naturally the case also for 
the conventional $\Delta v^{\text{BJ}}_{x \sigma}(\vr)$.
The calculation of the Slater part of the total potential is actually
still costly. For this part, we have previously provided accurate and numerically
simple approximations in 2D by considering the properties of the 
exchange hole.\cite{x1,ring}

\section{Results}

\subsection{Single-electron states of a harmonic oscillator}

First, we consider the single-electron non-interacting eigenstates in a 2D isotropic
harmonic oscillator. The external confining potential is given 
by $v_{0 \sigma}(r)=\omega^2 r^2/2$, where $\omega$ is the oscillator strength.
The use of this system is motivated by the frequent use of
such a confining potential when modeling 2D quantum dots~\cite{qd} (see
also below). The eigenstates, with the associated densities, 
are given by 
\bea
\phi_{nl}(r,\theta) &=& f_{nl}(r) \; e^{-r^2/2} \; e^{il\theta} \; , \\
\rho_{nl}(r) &=& |\phi_{nl}(r,\theta)|^2 = f^2_{nl}(r) \; e^{-r^2} \; ,
\eea
respectively. Here $r$ and $\theta$ are the usual radial and angular 2D polar coordinates.
$n = 0,1,2,...$ and $l = 0, \pm 1, \pm 2,...$ are the radial and
angular quantum numbers, and
$f_{nl}(r)$ is a radial function, related to Laguerre polynomials. For simplicity,
we do not consider the spin index in this analysis.
The corresponding contribution to the kinetic energy density is given by
\bea\label{th3}
\tau_{nl}(r) = |\nabla \phi_{nl}(r,\theta)|^2 &=&
\left[ \frac{df_{nl}(r)}{dr} - r f_{nl}(r) \right]^2 e^{-r^2} \nn \\
&+& \frac{l^2 \rho_{nl}(r) }{r^2} \;.
\eea
Following Eq.~(\ref{BJ2B}), we obtain
\be\label{BJBD}
\frac{\tau_{nl}(r)}{\rho_{nl}(r)} = \left[ \frac{df_{nl}(r)/dr}{f_{nl}(r)} 
- r \right]^2 + \frac{l^2}{r^2} \;.
\ee
Clearly, this contribution to $\Delta v_{x \sigma}(\vr)$ {\em diverges linearly} for $r \rightarrow \infty$,
after summing over all occupied states (all showing the same divergent
behavior), and after taking the square root of the sum. 
On the other hand, this divergent behavior is absent in our corrected
exchange potential in Eq.~(\ref{key}), as it cancels exactly with the second term
of $D_{\sigma}(\vr)$ in Eq.~(\ref{elf}).

\subsection{Many-electron quantum dots}

Next we focus on the same system as in the previous section but
consider $N$ interacting electrons. Within the effective-mass 
approximation for electrons in semiconducting host material such as
GaAs, this is the most common model for 2D quantum dots.~\cite{qd}

\begin{figure}
\includegraphics[width=0.80\columnwidth]{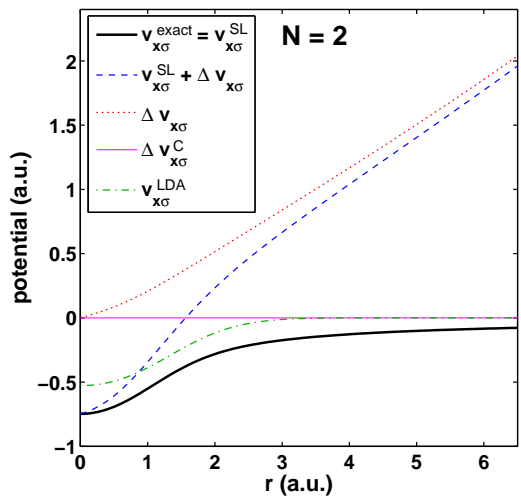}
\caption{(Color online) Exchange potentials for a two-electron quantum
  dot (singlet). The results are calculated from the analytic density.
$\Delta v_{x \sigma}$ and $\Delta v_{x \sigma}^{\text C}$ are defined in Eqs.~(\ref{BJ2B})
and (\ref{MBJ}), respectively.  
}
\label{fig1}
\end{figure}

First, we consider the singlet solution of $N=2$ and $\omega=1$ a.u., which
is known {\em analytically}.~\cite{taut} In this case, the exact exchange potential
is equal to the Slater potential (and minus half of the Hartree potential).
Therefore, any correction to the Slater potential should be zero.
For the kinetic-energy density we find
$\tau_\sigma(\vr)=(\nabla\rho_\sigma(\vr))^2/(4\rho_\sigma(\vr))$,
and the non-corrected Becke--Johnson-type approximation in
Eq.~(\ref{BJ2B}) becomes 
$\Delta v_{x\sigma}(\vr)=\sqrt{2}\nabla\rho_\sigma(\vr)/(3\pi\rho_\sigma(\vr))$,
where the spin density has an analytic expression
\begin{eqnarray} \label{density}                                                               
\rho_\sigma(r) & = & \frac{2}{\pi(\sqrt{2\pi}+3)}\Big\{e^{-r^2}(1+r^2/2)+              
\frac{1}{2}\sqrt{\pi}e^{-3r^2/2} \nonumber \\                            
& \times & \Big[I_0(r^2/2)+r^2 I_0(r^2/2)+r^2 I_1(r^2/2)\Big]\Big\},
\end{eqnarray}
deduced from the analytic wave function.~\cite{taut} Here
$I_0$ and $I_1$ are the zeroth and first-order modified Bessel 
functions of the first kind, respectively. Figure~\ref{fig1}
shows that $\Delta v_{x\sigma}(\vr)$ (dotted line) has clear divergent
behavior, making the total exchange potential (dashed line) largely different
from the exact result (thick solid line), even in the small-$r$ regime.
It should be noted that also the LDA exchange potential considerably
deviates from the exact result. On the other hand, our corrected scheme
in Eq.~(\ref{key}) yields the exact result: 
$\Delta v^{\rm C}_{x\sigma}(\vr)=0$ as discussed above.

The divergent behavior of
$\Delta v_{x \sigma}(r)$ may be easily extracted from the interacting density
given in Eq.~(\ref{density}). For this, one should use the asymptotic expansion of
the modified Bessel functions: $I_{\nu}(z \rightarrow \infty)
\rightarrow e^z/\sqrt{2 \pi z}+ ...$ (Ref.~\onlinecite{AS}).
Using this in Eq.~(\ref{density}) leads to 
$\rho_{\sigma}(r) \rightarrow r^2e^{-r^{2}}/\pi(3+\sqrt{2 \pi})$, and
$\nabla \rho_{\sigma}(r) \rightarrow -2r^3e^{-r^{2}}/\pi(3+\sqrt{2 \pi})$
in the asymptotic region ($r \rightarrow \infty$). Replacing everything,
we obtain that $\Delta v_{x{\sigma}}(r \rightarrow \infty) \rightarrow 2\sqrt{2}r/(3 \pi) \simeq 0.300~r$
in agreement with Fig.~{\ref{fig1}}. It is interesting
to note that the leading term to the divergent contribution obtained for the
interacting system agrees exactly with the non-interacting estimate given above in
Eq.~(\ref{BJBD}). The reason is that the density is
rather small in the asymptotic region, and then the modifications 
introduced by interactions to the bare 2D harmonic potential are also rather small. 
As a consequence, the interacting densities approach the non-interacting limit asymptotically.

\begin{figure}
\includegraphics[width=0.80\columnwidth]{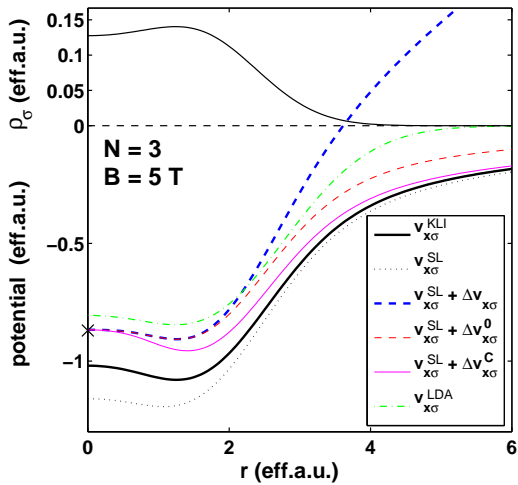}
\caption{(Color online) Exchange potentials for a fully spin-polarized 
three-electron quantum dot in external magnetic field of $B=5$ T.  
Terms with $\Delta v_{x \sigma}^{\text C}$ and $\Delta v_{x \sigma}^{\text 0}$
refer to the corrected
approximation [Eq.~(\ref{key})], with and without the explicit current term
in Eq.~(\ref{elf}), respectively.  
The scales for the vertical axis referring to the potentials and density (negative and positive values, respectively)
are different.
}
\label{fig2}
\end{figure}

\begin{figure}
\includegraphics[width=0.80\columnwidth]{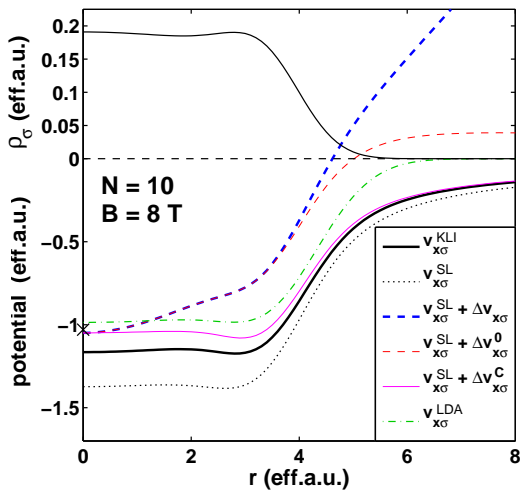}
\caption{(Color online) Same as Fig.~\ref{fig2} but for ten electrons at
$B=8$ T.
}
\label{fig3}
\end{figure}

In Figs.~\ref{fig2} and \ref{fig3}
we show the exchange potentials $v_{x\sigma}$ for quantum dots
containing three and ten interacting electrons, respectively. 
In both cases the oscillator strength is 
$\omega=0.4217$ in {\em effective} atomic
units (eff.a.u.) with the material parameters for GaAs: 
$m^*=0.067\,m_e$ and $\epsilon=12.4\,\epsilon_0$.
Here we have performed self-consistent KLI calculations by applying
the {\tt octopus} DFT code,~\cite{octopus}
 and use the resulting KS orbitals as
inputs in the approximations given above. Both
dots are exposed to an external, perpendicular magnetic field 
with a sufficient strength to achieve full spin polarization $(S=N/2)$ and
occupation to eigenstates of consecutive angular momenta from 
$l=0$ to $l=-N+1$. The resulting ``maximum-density droplet''~\cite{mdd} has a 
smooth electron density (see the upper panels of 
Figs.~\ref{fig2} and \ref{fig3}). The corresponding exchange
potentials calculated with different approximations have several 
interesting features listed below.

\begin{itemize}

\item Similarly to the previous examples, the straightforward BJ 
extension, i.e., Slater potential combined with $\Delta v_{x\sigma}$
in Eq.~(\ref{BJ2B}), shows divergent behavior.

\item Our approximation $\Delta v_{x \sigma}^{\text C}(\vr)$ in
  Eq.~(\ref{key}) leads to
  good overall agreement with the KLI potential, especially at large
  $r$. The agreement improves as a function of $N$. 

\item The LDA exchange potential is overall worse than our approximation
-- not only in the asymptotic region showing the obvious exponential
decay -- but also close to the center of the system. 

\item The current-dependent term introduced in Eq.~(\ref{current}),
which enforces the gauge invariance,
is crucial in finding the correct behavior. Without this term
($\Delta v_{x \sigma}^{\text 0}$) 
the performance is very poor at high magnetic fields
(see in particular Fig.~\ref{fig3} with $B=8$ T).

\item Close to the center of the maximum-density droplet, which 
locally resembles the 2DEG with the electron localization function
$\approx 1/2$ (Ref.~\onlinecite{elf3}), the present approximation 
satisfies with a good precision the well-known exact relation for the homogeneous
2DEG between the 
exchange potential and the Slater potential: 
$v_{x\sigma} = (3/4) v^{\rm SL}_{x\sigma}$ (see the crosses in 
Figs.~\ref{fig2} and \ref{fig3}). For the calculation of the cross positions,
the numerical value of $v_{x \sigma}^{{\rm SL}}(r=0)$ has been multiplied by 3/4. Note that this relation is not
well satisfied between $v_{x \sigma}^{{\rm LDA}}(r=0)$ and $v_{x \sigma}^{{\rm SL}}(r=0)$,
because the quantum-dot system is finite, and globally not homogeneous.

\end{itemize}

\begin{figure}
\includegraphics[width=0.80\columnwidth]{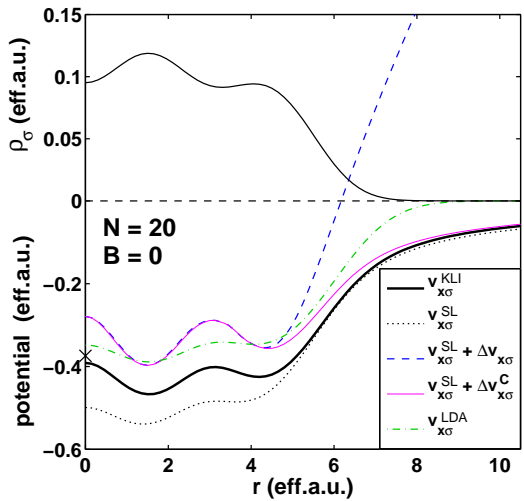}
\caption{(Color online) Exchange potentials for a closed-shell quantum
  dot with 20 electrons.
}
\label{fig4}
\end{figure}

Figure~\ref{fig4}
shows the exchange potentials for a closed-shell quantum dot with 
20 electrons.
Again, the straightforward BJ extension to 2D diverges,
whereas the cured approximation qualitatively agrees with the KLI 
solution. It should be noted that the shell structure is better 
reproduced by the present approximation than by the LDA.
In this system, the  $(3/4)$-relation is not well satisfied (see the cross
in Fig.~\ref{fig4}), since the system properties in the core region
are far from those of the homogeneous 2DEG, i.e., the electron localization
function is varying.~\cite{elf3}
Also note that for this closed-shell system, the spin-dependent 
paramagnetic current density
vanishes, and then $\Delta v_{x \sigma}^{\text C}(\vr) = \Delta v_{x \sigma}^{0}(\vr)$.

Finally, we point out that the present scheme, in particular
the modifications introduced in Eqs.~(\ref{current}) and (\ref{D}),
can be considered and reworked also in 3D. 
This has been shown to lead to an approximation
for the 3D exchange potential that gives 
good results in systems possessing orbital currents 
(e.g., open-shell systems), in the presence of external
magnetic or electric fields, or in ``non-atomic'' systems
such as the Hooke's atom, i.e., the 3D counterpart of the
harmonic quantum dot.~\cite{jcp}

\section{Conclusions}

To achieve an accurate and computationally convenient approximation 
to the exchange potential in two dimensions,
we have applied and extended the formalism of Becke and Johnson
commonly used in three dimensions. We have found that direct
extension of the formalism leads to divergent behavior of the exchange
potential in the most standard two-dimensional applications. 
We have corrected the scheme by introducing
a current-dependent term ensuring the gauge invariance and
a density-gradient term capturing the disappearance of the
correction to the Slater contribution
for a generic single-particle state. The resulting approximation 
for the exchange potential agrees well with the Krieger-Li-Iafrate
potential in various quantum dots up to high magnetic fields and
thus to high current densities. The correct asymptotic behavior is
recovered and the shell-structure is well described. 
Similar strategy is expected to
lead to good results also in three-dimensional applications in 
external magnetic and/or electric fields.

\begin{acknowledgments}
This work was supported by the Deutsche 
Forschungsgemeinschaft and the Academy of Finland.
C.R.P. was supported by the European Community through a Marie Curie
IIF (Grant No. MIF1-CT-2006-040222) and CONICET of
Argentina through Grant No. PIP 5254.
\end{acknowledgments}

\end{document}